\documentclass[preprint,authoryear,12pt]{elsarticle}

 \usepackage{graphicx}
\usepackage{amsmath} 
\usepackage{amssymb}

\journal{Journal of Theoretical Biology}

\begin{document}

\begin{frontmatter}



\title{Life history and mating systems select for male biased parasitism mediated through natural selection and ecological feedbacks}


\author[1]{Flora S. Bacelar}
\author[2]{Andrew White}
\author[3]{Mike Boots }
\address[1]{IFISC(CSIC-UIB) Instituto de F{\'\i}sica Interdisicplinar y Sistemas
Complejos, Campus Universitat Illes Balears, E-07122 Palma de Mallorca, Spain (florabacelar@ifisc.uib-csic.es)}
\address[2]{Department of Mathematics and the Maxwell Institute for Mathematical Sciences, Heriot-Watt University, Edinburgh, EH14 4AS, UK. (A.R.White@hw.ac.uk)}
\address[3]{Department of Animal and Plant Sciences, University of Sheffield, Sheffield, S10 2TN, UK. (m.boots@shef.ac.uk)}

\begin{abstract}
Males are often the "sicker" sex with male biased parasitism found in a taxonomically diverse range of species. There is considerable interest in the processes that could underlie the evolution of sex-biased parasitism. Mating system differences along with differences in lifespan may play a key role. We examine whether these factors are likely to lead to male-biased parasitism through natural selection taking into account the critical role that ecological feedbacks play in the evolution of defence. We use a host-parasite model with two-sexes and the techniques of adaptive dynamics to investigate how mating system and sexual differences in competitive ability and longevity can select for a bias in the rates of parasitism. Male-biased parasitism is selected for when males have a shorter average lifespan or when males are subject to greater competition for resources. Male-biased parasitism evolves as a consequence of sexual differences in life history that produce a greater proportion of susceptible females than males and therefore reduce the cost of avoiding parasitism in males. Different mating systems such as monogamy, polygamy or polyandry did not produce a bias in parasitism through these ecological feedbacks but may accentuate an existing bias.
\end{abstract}


\end{frontmatter}


\section{Introduction}
Across a diverse range of taxa there is often a sex bias in parasitism rates, with males commonly the "sicker" sex \citep{Zuk1990,Zuk2009}. Although male-biased parasitism is by far the most commonly reported sex-biased parasitism \citep{Poulin1996,Schalk1997,Perkins2003,Ferrari2004}, higher rates of parasitism have also been reported for females in avian hosts \citep{McCurdy1998}. It remains unclear what processes can account for sex-biased parasitism and in particular the higher prevalence of disease in males. One possibility is that males exhibit different behaviour that leads to greater exposure (i.e. larger home ranges or more risk of infection for any given exposure because of damage caused by fighting) \citep{Bundy1988,Restif2010}. Bias may also result from underlying differences in life-history characteristics between males and females \citep{Moore2002} including the idea that the larger physical size and growth rates of males make them a more accessible and attractive target for parasites \citep{Moore2002}. There is also clear evidence of a physiological basis for differences in susceptibility with for example androgenic hormones in males (testosterone in vertebrates), acting to depress the immune system \citep{Moore2002,Folstad1992,Alexander1988}.

Beyond physiological mechanisms, it has been proposed that life-history theory could explain immune differences from an adaptive point of view in relation to sex-specific reproductive strategies. In particular it has been argued that the reduced investment in susceptibility is due to trade-offs between male mating effort and immune defense. In this scenario as the strength of sexual selection on males increases, the magnitude of the sex differences in immunocompetence will increase. In essence the argument is that a reduced immune systems may be the unavoidable price of being male due to sexual selection \citep{Zuk1990,Zuk2009}. Following on from this a polygamous mating systems should lead to greater differences in male biased parasitism and that perhaps under polyandry females should be more susceptible \citep{Zuk1990,Zuk2009}. The basic assumptions of these ideas have recently been examined theoretically by \citet{Stoehr2006} who determine optimal allocation of resources between immunity, survival and reproduction in males and females, under varying levels of sexual selection. This work has given a strong theoretical underpinning to ideas that sexual selection can explain male-biased parasitism \citep{Stoehr2006,Zuk2009}. In addition, \citet{Moore2002} carried out a meta-analysis using two measures of the strength of sexual selection-mating system and sexual size dimorphism-that showed that sexual selection was associated with sex differences in parasitism.

There is a large body of theoretical work that has emphasised the importance of ecological feedbacks to the evolution of host defence to infectious disease (see \citet{BOOTS2009} for a review). It is clear from this theory that host life-history is critical to level of defence that evolves. In particular, it is often, although not always the case, that increased resistance to parasites is more likely to evolve for long-lived hosts \citep{Miller2007b}. Differences between males and females in terms of their life-histories may therefore be enough to explain the evolution of different levels of investment in defence through natural rather than sexual selection. In particular differences in ecological feedbacks between males and females due to differences in their life-histories may underpin the evolution in reduced investment by males in defence and therefore result in higher transmission of infection. Furthermore, the ecological feedbacks due to monogamous and polygamous mating systems have different effects on the evolution of male and female investment in defence. In particular, different mating systems cause different patterns in the way in which densities of males and females feedback into the evolutionary dynamics.

A recent model \citep{Restif2010} has shown the importance of epidemiological feedbacks in determining male-biased parasitism through natural selection. \citet{Restif2010} focus on trade-offs in the disease characteristics such that differences in exposure may lead to differential investment in resistance. We develop a continuous time host-parasite system that represents two-sexes and investigates how the mating system and differences in competitive ability and longevity between sexes influences the level of resistance to infection that evolves. As such we are examining the effects of mating system and host life-history under natural rather than sexual selection. Crucially we include epidemiological dynamics and focus therefore on how ecological feedbacks affect the evolutionary process. We find that differences in lifespan are enough to explain male biased parasitism. Differences in the mating system act only to accentuate an existing bias. Our work further emphasises the importance of including epidemiological feedbacks when studying the evolution of defence.

\section{Methods}

The underlying host-parasite framework is based on the classical approaches for modelling the population dynamics of directly transmitted microparasites (see Anderson1981) and which have been successfully extended to understand the evolution of host resistance \citep{BOOTS1999a,BOOTS2009}. This framework is modified following the techniques of \citet{LINDSROM1998} and \citet{MILLER2007a}, to represent a two-sex host parasite model that considers the dynamics of males and females separately. This is achieved by representing births as the harmonic mean function proposed by \citet{CASWELL1986}, that depends on the densities of the two sexes and declines to zero in the absence of either sex. This function can also be modified to approximate different mating systems (monogamy, polygyny and polyandry). The theoretical framework is therefore represented by the following system of nonlinear ordinary differential equations for the densities of susceptible, $S$, and infected, $I$, males and females, represented by the subscripts $m$ and $f$ respectively.

\begin{eqnarray}
   \frac{dS_{m}}{dt} &=& \frac{1}{2}B(S_{m},S_{f})(1-q_{m}H)-b_{m}S_{m}-\beta _{m} S_{m}(I_{f}+I_{m})\nonumber\\
   \frac{dI_{m}}{dt} &=&   \beta _{m}S_{m}(I_{f}+I_{m})-(b_{m}+\alpha)I_{m} \\
   \frac{dS_{f}}{dt} &=& \frac{1}{2}B(S_{m},S_{f})(1-q_{f}H)-b_{f}S_{f}-\beta _{f} S_{f}(I_{f}+I_{m})\nonumber\\
  \frac{ dI_{f}}{dt} &=&  \beta _{f}S_{f}(I_{f}+I_{m})-(b_{f}+\alpha)I_{f}\nonumber\\\nonumber
 \end{eqnarray}

Where $H= S_{f}+S_{m}+I_{f}+I_{m}$  is the total host density. Births are divided equally between males and females according to the harmonic birth function, $B(S_{m},S_{f})$  , which describes the dependency of the birth rate on the density of either sex and mating strategy. The birth rate is modified due to density-dependent competition for resources with the parameter $q$, and the population has a natural death rate, $b$. Infection can occur through contact between susceptible and infected individuals with transmission coefficient, $\beta$, and the disease induces an additional mortality while infected at rate $\alpha$. (The subscripts on some parameters distinguishes between male and female specific parameters.) The harmonic birth function, $B(S_{m},S_{f})$  , is derived from \citep{CASWELL1986}, and takes the following form.
 \begin{equation}\label{birthfunction}
   B(S_{m},S_{f})= \frac{c_{m}S_{m}c_{f}S_{f}}{S_{m}+\frac{S_{f}}{h}}
\end{equation}

Here, $c_m$ and $c_f$ represent the contribution that males and females make to the birth rate and $h$ represents harem size and can be manipulated to represent different mating systems. When $h>1$ it represents a polygenic mating system (births are maximised when females exceed males), when $h<1$ it represents polyandry (births are maximised when males exceed females) and when $h=1$ it represents monogamy (births are maximised when males and females are equally abundant) \citep{CASWELL1986}.

To examine the evolution of parasite resistance we follow the techniques of adaptive dynamics \citep{GERIZ1998,BOOTS2009}. We assume a 'mutant' strain of host can occur at low density and attempt to invade the established 'resident' strain which is at its equilibrium density. The mutant male host strain differs from the resident strain in terms of its transmission coefficient $\tilde{\beta_m}$   compared to $\beta_m$  for the resident (a similar difference can occur for the female transmission coefficient and we will use `` $ \tilde{}$ '' to represent the mutant parameters). In line with previous studies into the evolution of host resistance it is assumed and that a benefit in terms of increased resistance to infection is bought at a cost in terms of a reduced birth rate \citep{BOOTS1999a}. For this study we impose the trade-off $c_m=g(\beta _m)$  and $c_f=g(\beta _f)$ . The trade-off is defined by
\begin{equation}\label{trade-off}
    c_m = c_{max}- \left( (c_{max}-c_{min})\frac{\left(1-\frac{\beta m-\beta min}{\beta max-\beta min}\right)}{\left(1+\gamma \frac{\beta m-\beta min}{\beta max-\beta min}\right)}\right)
\end{equation}

Which is a smooth curve between the minimum and maximum values of the birth and transmission rates and in which the parameter $\gamma$  controls the curvature (and therefore cost structure) of the trade-off. (We will discuss other possible trade-offs later.)

The fitness is the long-term exponential growth rate of a phenotype in a given environment. We initially consider the situation where the female parameters are fixed and we allow the male parameter $\beta _m$ (and $c_m$ via the trade-off) to evolve. A proxy for the fitness, $R$, can be calculated as the determinant of the jacobian matrix, $J$, at the steady state $(S_m,S_f,I_m,I_f,\tilde{S}_m,\tilde{I}_m)=(S_m^*,S_f^*,I_m^*,I_f^*,0,0)$ \citep{MILLER2005} where

\begin{equation}\label{mutantjacobian}
    J=\left(
        \begin{array}{cc}
          \frac{\partial {\tilde{\dot{S}}_{m}}}{\partial {\tilde{S}_{m}}}  & \frac{\partial {\tilde{\dot{S}}_{m}}}{\partial {\tilde{I}_{m}}} \\
         \frac{\partial {\tilde{\dot{I}}_{m}}}{\partial {\tilde{S}_{m}}} & \frac{\partial {\tilde{\dot{I}}_{m}}}{\partial {\tilde{I}_{m}}} \\
        \end{array}
      \right)
\end{equation}

and therefore, $R$, can be represented by the following expression
\begin{eqnarray}\label{fitness}
    R &=&-b_{m}-(I_{f}^{*}+I_{m}^{*})\tilde{\beta}_{m} +\frac{c_{f}a_{m}g(\tilde{\beta_m})S_{f}^{*}(1-q_{m}H^*)}{2(\frac{S_{f}^{*}}{h}+S_{m}^{*})}\\
    H^*&=&S^*_m+S_f^*+I_m^*+I_f^*\nonumber
 \end{eqnarray}
The fitness proxy, $R$, can be used to determine the position of evolutionary singular points and the evolutionary behaviour at the singular point. Evolutionary singular points are determined when the fitness gradient
$\frac{\partial R}{\partial\tilde{\beta}_m}\Bigr\rvert_{\beta_m=\tilde{\beta}_m}=0$ which equates to solving the following expression
\begin{eqnarray}\label{EP}
 -I_f - I_m +\frac{1}{2(\frac{S_f}{h}+S_m)}a_m c_f S_f (1-q_m H^*)g'(\widetilde{\beta}_{m})=0
  \end{eqnarray}

The evolutionary behaviour at the singular point is determined by analysing the second order partial derivatives of $R$ with respect to the mutant and resident parameters \citep{GERIZ1998}. Previous studies have assessed how the trade-off cost structure or underlying population dynamics can influence the evolutionary behaviour and also induce evolutionary branching leading to diversity in host strategies \citep{BOOTS1999a,BOOTS2009}. The focus here is to examine whether different levels of resistance can evolve between males and females and therefore to allow us to concentrate on this issue we ensure that the underlying population dynamics are point equilibrium and that the trade-off has sufficiently accelerating costs that the singular point is an evolutionary stable attractor. We can then assess how the position of the singular point changes as other life history parameters are varied.

To examine the coevolution of male and female resistance properties we determine the female fitness function (which depends on the evolving parameter $\beta_f$  and $c_f$ via the trade-off). The male singular points are plotted against $\beta_f$  (for a fixed female strategy using the method outlined above) and the female singular points (against a fixed male strategy) are plotted against $\beta_m$ . The intersection of these lines produces a coevolutionary attracting singular point (\citet{Restif2003}, note that again the trade-off is chosen to ensure the population dynamics exhibit a point equilibrium and that the singular point is a coevolutionary stable attractor). Using this method it is possible to determine how the coevolutionary singular point varies with changes in underlying life history parameters.

\section{Results}
\begin{figure}[ht]
  \begin{center}
  $\begin{array}{cc}
  \multicolumn{1}{l}{\mbox{\bf a)}} &  \multicolumn{1}{l}{\mbox{\bf b)}} \\ [-1cm]
   \includegraphics[width=5.5cm]{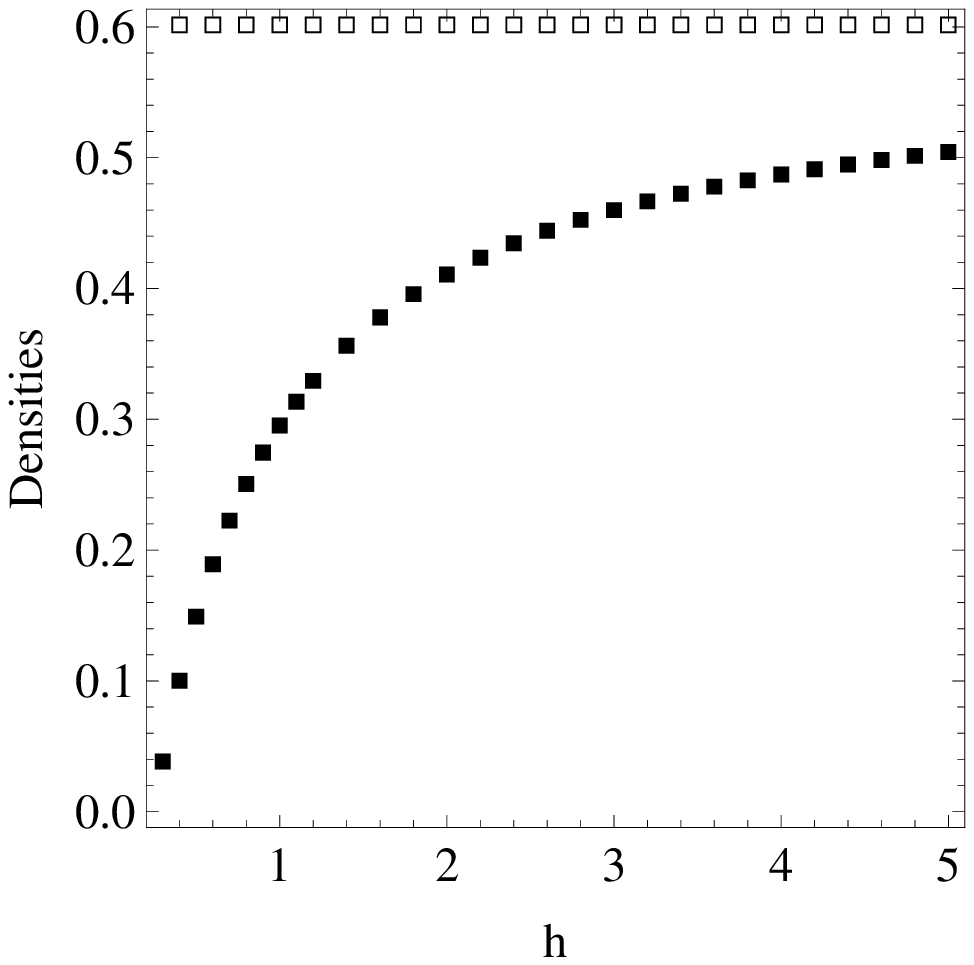} & \includegraphics[width=5.5cm]{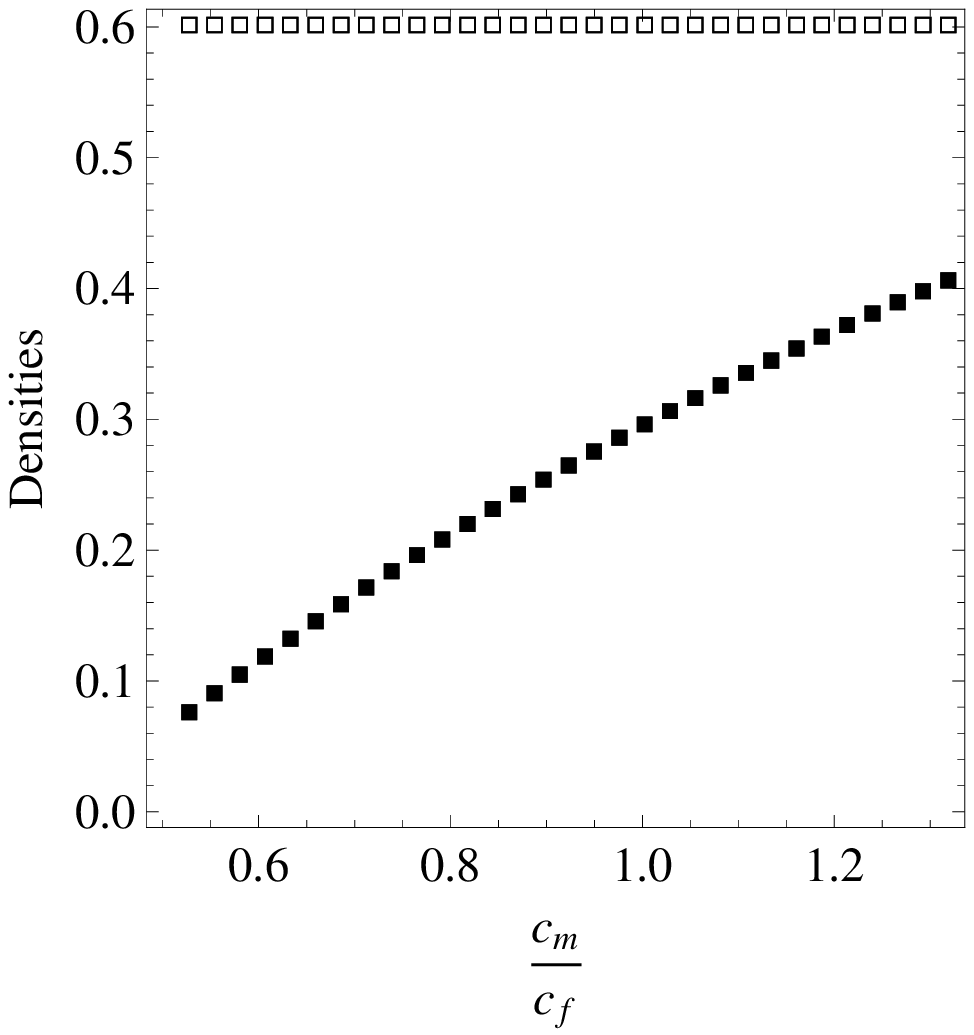}\\
    \multicolumn{1}{l}{\mbox{\bf c)}} &  \multicolumn{1}{l}{\mbox{\bf d)}} \\ [-1cm]
    \includegraphics[width=5.5cm]{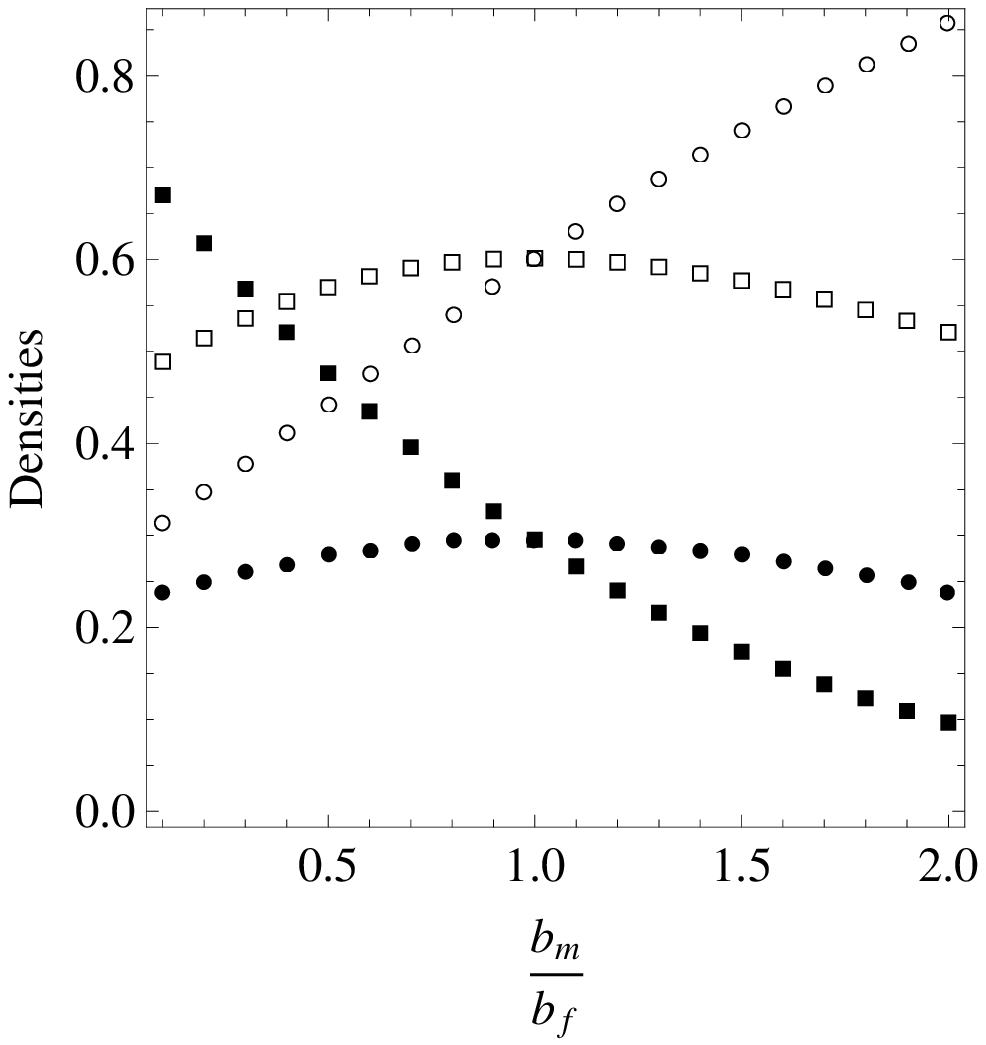} &  \includegraphics[width=5.5cm]{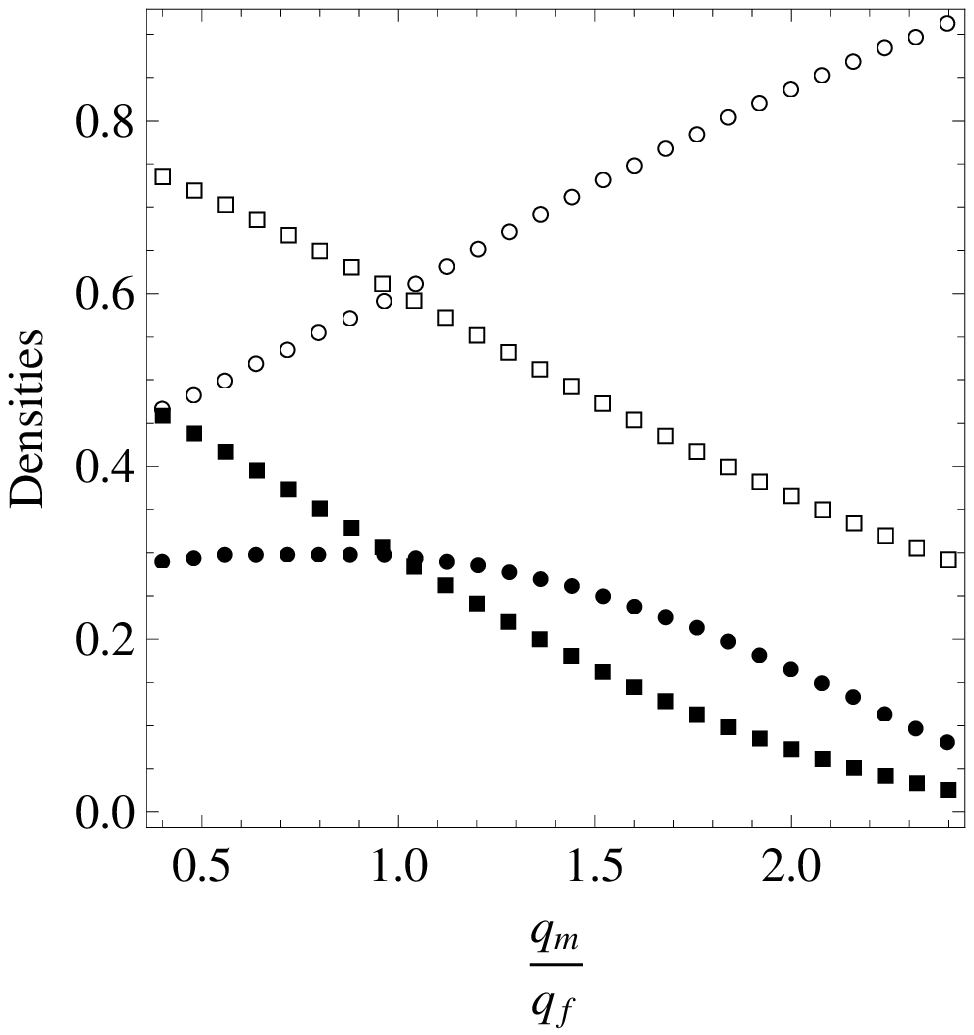}
  \end{array}$
  \caption{Changes in the density of the different classes as other parameters are varied where ``$\circ$" represents $S_f$, ''$\bullet$" represents $I_f$, ``$\scriptscriptstyle\square$" represents $S_m$ and ``$\scriptscriptstyle\blacksquare$" represents $I_m$. Parameter values (unless varied in the figure) are $c_m=c_f=3.79$, $q_m=q_f=0.25$, $b_m=b_f=1$, $\alpha_m=\alpha_f=1$, $\beta_m=\beta_f=1.6625$, $h=1$. In a) $h$ a is varied,  b) $c_m$ is varied, c) $b_m$ is varied and d) $q_m$ is varied. In a) and b) $S_f =S_m$ and $I_f=I_m$ so the results are just shown for males.}\label{Fig1density}
  \end{center}
\end{figure}

The evolutionary behaviour is dependent on feedbacks that arise in the ecological dynamics and therefore it is useful to first understand how changes in life history parameters will affect the equilibrium density of the different classes in the model system (Figure \ref{Fig1density}). Increases in the harem size, $h$, or the overall birth rate leads to an increase the total density of males and females (equally since the sex ratio is 50:50). If the male death rate is reduced (relative to the female death rate) then there is an increase in the overall density of males (through an increase in infected males) and a decrease in female density. As the male death rate is increased then there is an increase in the overall density of females (through an increase in susceptible females) and a decrease in male density. Note also that the prevalence of infection decreases as the male death rate increases (relative to the female death rate). When there is a reduction in the competition parameter for males there is an increase in male density through increases in susceptible and infected males and female density in both susceptible and infected classes is reduced. When there is an increase in the competition parameter for males then overall male density and female infected density decreases while female susceptible density increases. The overall susceptible density remains constant as the competition parameter is varied but the proportion of susceptible males/susceptible females decreases as the competition parameter for males is increased. Also, the prevalence of infection remains relatively constant when the male competition parameter is less than the female parameter but the prevalence decreases when the male competition parameter is greater than the female parameter.

\subsection{Evolving male characteristics.}
\begin{figure}[ht]
  \begin{center}
  $\begin{array}{ccc}
   \multicolumn{1}{l}{\hspace{-0.1cm}\mbox{\bf a)}} &  \multicolumn{1}{l}{\hspace{-0.2cm}\mbox{\bf b)}} &  \multicolumn{1}{l}{\hspace{-0.2cm}\mbox{\bf c)}} \\ [-1cm]
     \includegraphics[width=4.8cm]{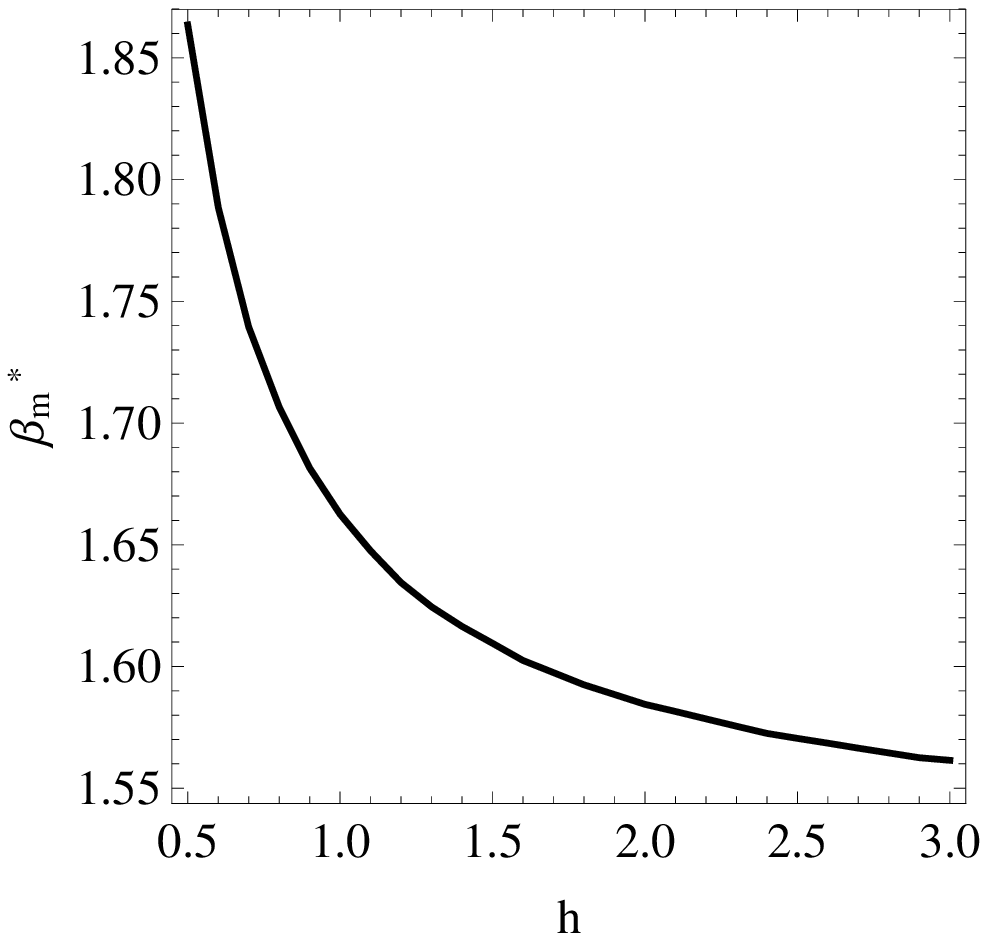} & \includegraphics[width=4.7cm]{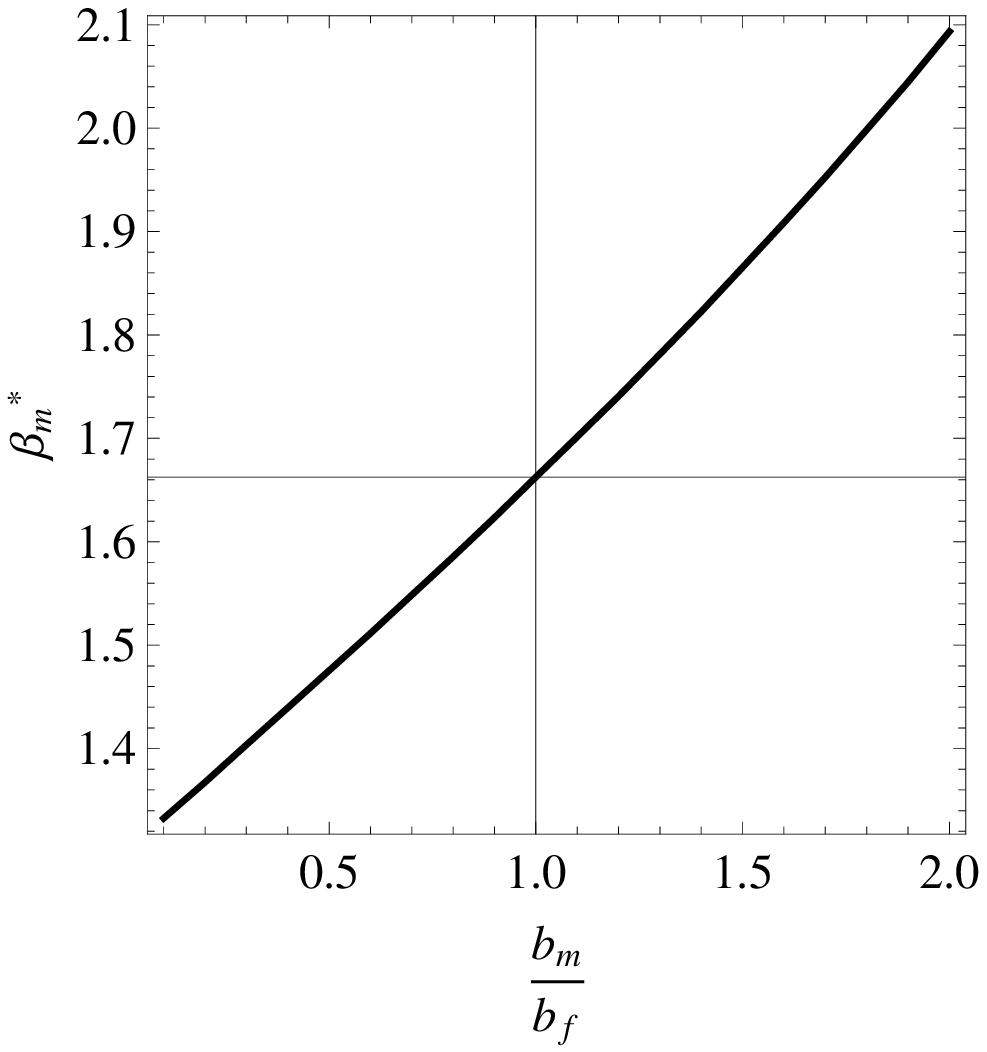} & \includegraphics[width=4.7cm]{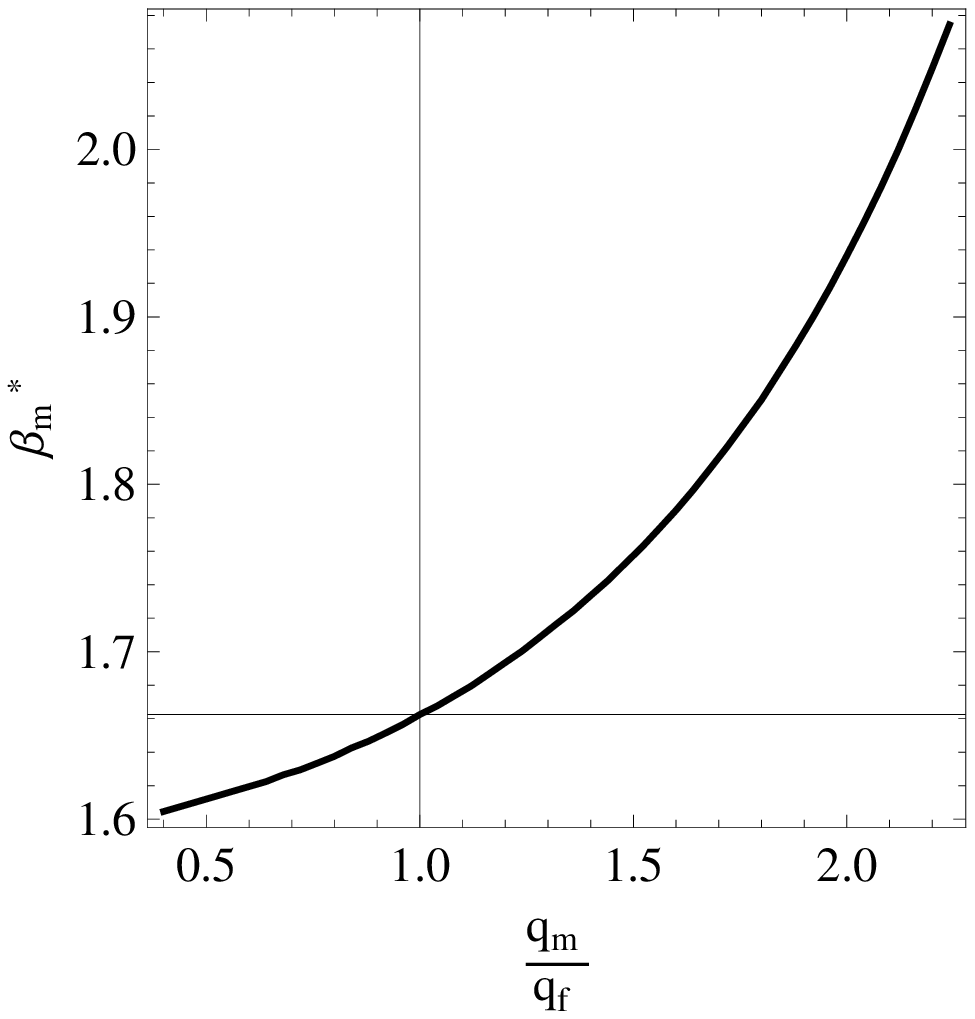}
   \end{array}
  $
   \caption{Change in the singular value of disease transmission for males, $\beta_m^*$, when males can evolve ( and females do not evolve). Parameters values (unless varied in the figure) are $c_f=3.79$, $q_m=q_f=0.25$, $b_m=b_f=1$, $\alpha_m=\alpha_f=1$, $\beta_f=1.6625$, $h=1$. The parameters for the trade-off (equation \ref{trade-off}) are $\beta_{min}=1$, $\beta_{max}=20$, $c_{min}=2$, $c_{max}=5$ and $\gamma=40$. In a) h is varied, b) $b_m$ is varied and c) $q_m$ is varied.}\label{Fig2cbeta-h}
  \end{center}
\end{figure}

The results when the male characteristics are allowed to evolve against fixed female parameters are shown in figure \ref{Fig2cbeta-h}. As harem size decreases the level of resistance to disease in males decreases (the singular value of $\beta_m$ increases). Decreased disease resistance in males also evolves as the male death rate increases and the male competition parameter increases. This decrease in disease resistance in males is a response to decreased levels of prevalence of infection (see figure \ref{Fig1density}) which reduces the need to avoid infection (as individuals are less likely to become infected). Male-biased parasitism is therefore evident under polyandrous mating systems and when males have a higher death rate or suffer more severe competition than females.

\subsection{Coevolving male and female characteristics}

\begin{figure}[ht]
  \begin{center}
 $ \begin{array}{cc}
  \multicolumn{1}{l}{\mbox{\bf a)}} &  \multicolumn{1}{l}{\mbox{\bf b)}} \\ [-1cm]
    \includegraphics[width=6.5cm]{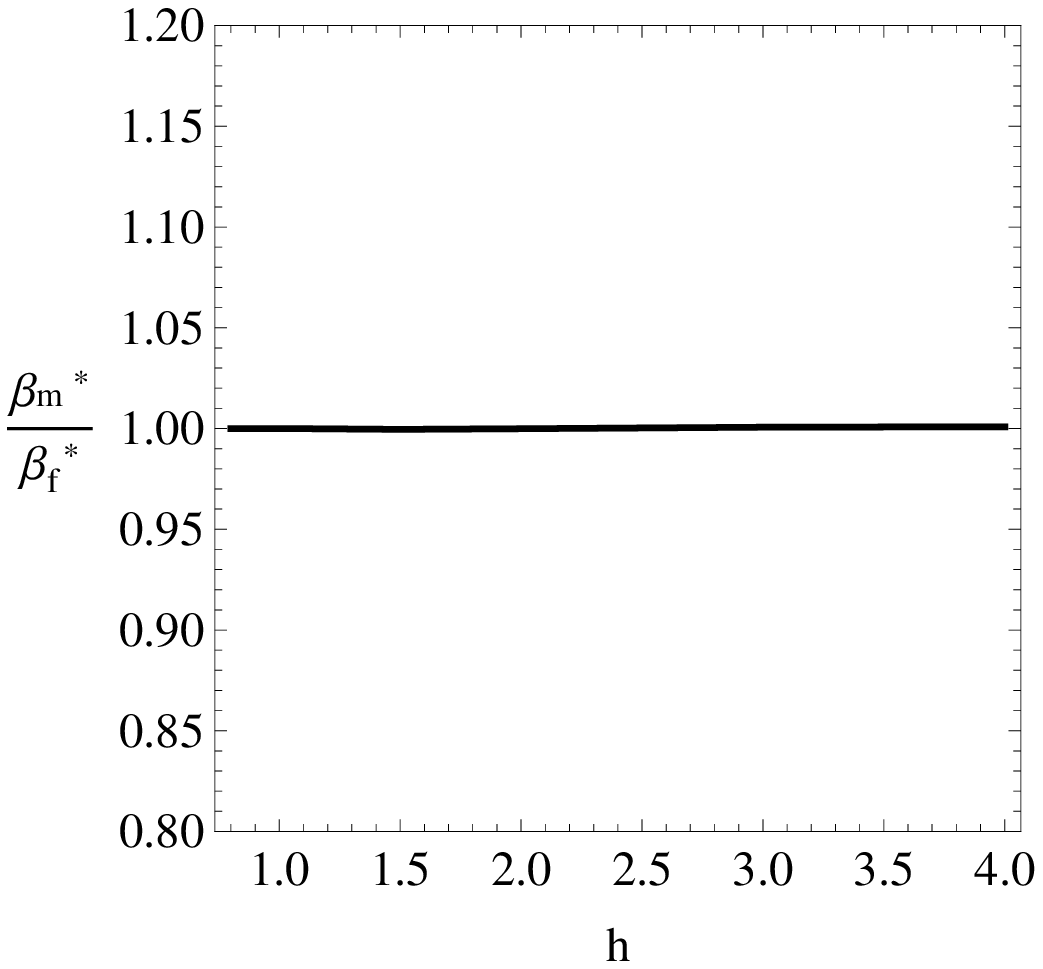} & \includegraphics[width=6.5cm]{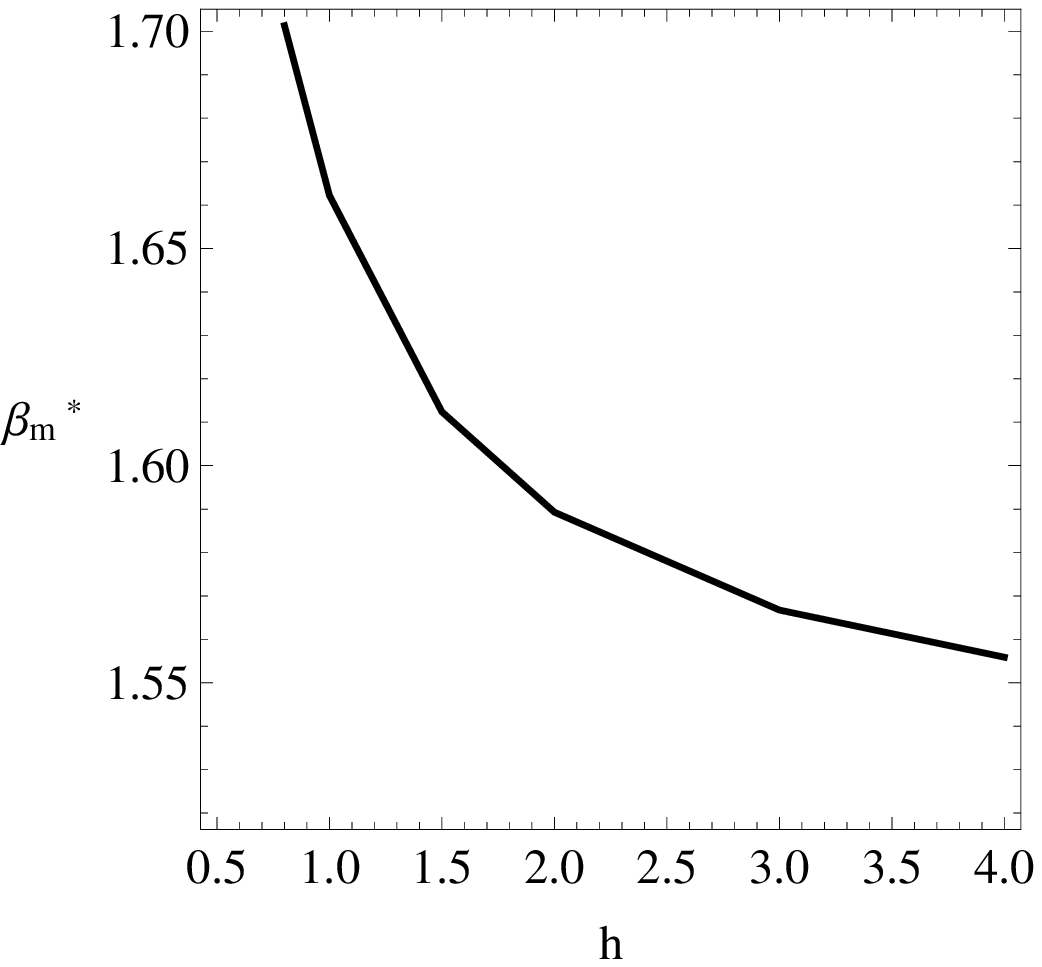}
  \end{array}$
  \caption{Change in the coevolutionary singular value of disease transmission, $\beta_m^*$ and $\beta_f^*$ plotted against changes in harem size. Parameter values are $q_m=q_f=0.25$, $b_m=b_f=1$, $\alpha_m=\alpha_f=1$. The parameters for the trade-off (equation \ref{trade-off}) are as in Fig.\ref{Fig2cbeta-h}. a) shows the relative values of $\beta_m$ and $\beta_f$ at the coevolutionary singular point and b) shows the actual value of $\beta_m^{*}$ at the coevolutionary singular point (note the value of $\beta_f^{*}$ is identical here).}\label{Fig3coevolpoint_h}
  \end{center}
\end{figure}

When both male and female characteristics are allowed to evolve variation to the harem size does not produce a bias between male and female infection rates (figure \ref{Fig3coevolpoint_h}a). Both males and females will evolve increased levels of resistance as harem size increases (figure \ref{Fig3coevolpoint_h}b) in response to the associated increases in prevalence.

\begin{figure}[ht]
  \begin{center}
  $\begin{array}{cc}
   \multicolumn{1}{l}{\mbox{\bf a)}} &  \multicolumn{1}{l}{\mbox{\bf b)}} \\ [-1cm]
    \includegraphics[width=6.5cm]{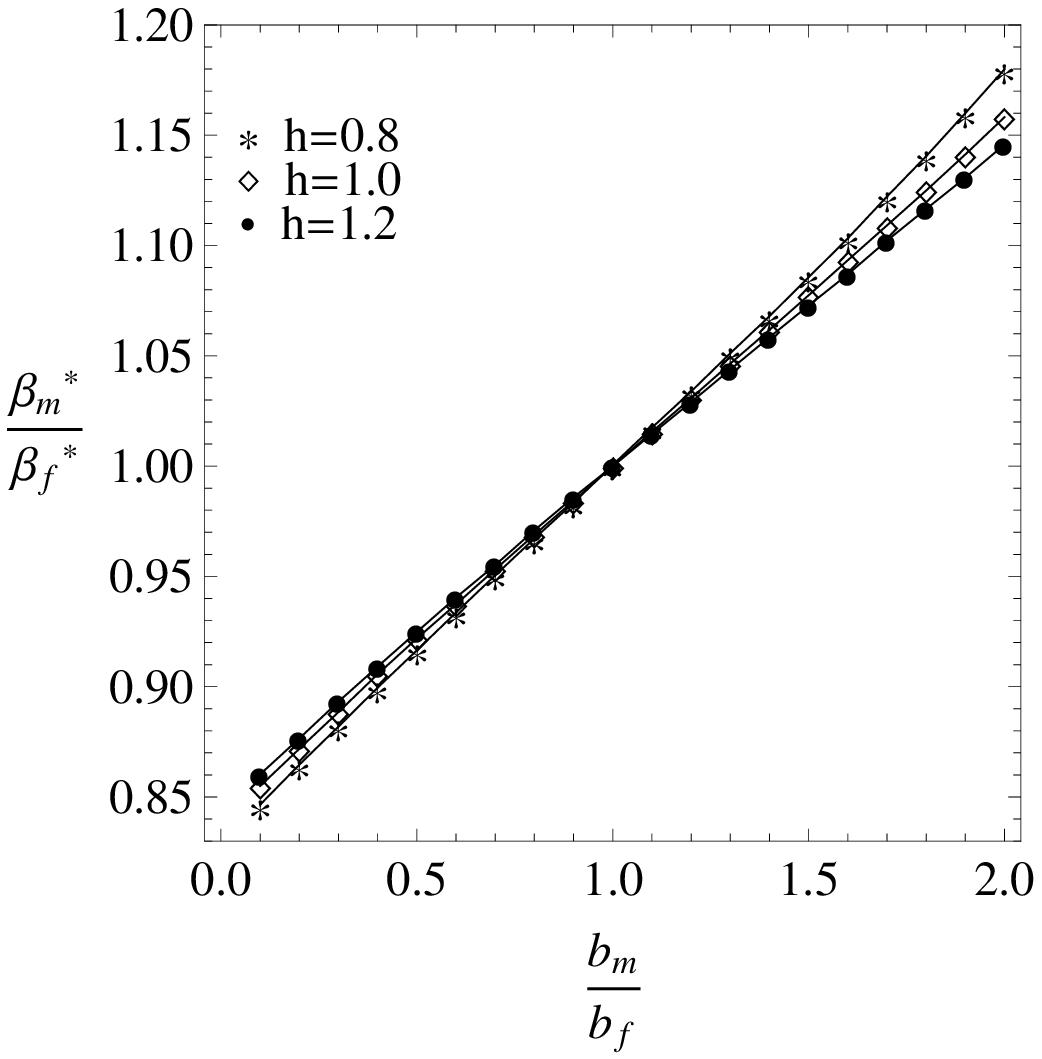} &  \includegraphics[width=6.5cm]{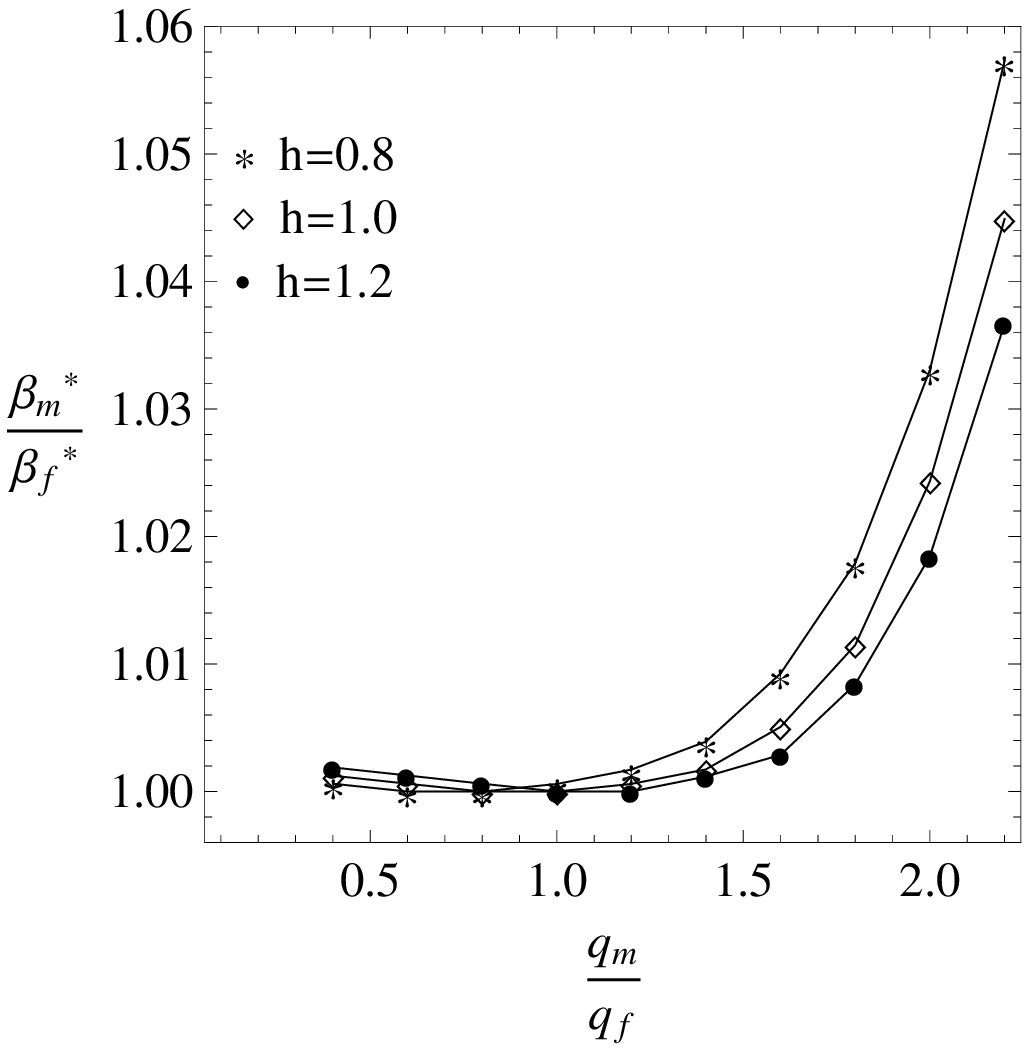}
  \end{array}$
   \caption{Change in the coevolutionary singular value of disease transmission $\beta_m^*$ and $\beta_f^*$ plotted against changes in a) male birth rate and b) male susceptibility to crowding. Parameter values (when not varied in the figure) are $b_m=bf=1$, $q_m=q_f=0.25$, $\alpha_m=\alpha_f=1$, $\beta_m=\beta_f=1.6625$. The parameters for the trade-off (equation \ref{trade-off}) are as in Fig.\ref{Fig2cbeta-h}.}\label{Fig4coevolpoint_qmbm}
  \end{center}
\end{figure}

When the male death rate exceeds that of females then male biased parasitism can result from coevolution (figure \ref{Fig4coevolpoint_qmbm}a). Here the increased death rate for males means they have on average a shorter lifespan and so increases the possibility of dying from natural causes before becoming infected. This is reflected in the fact that the proportion of female/male susceptibles increases as the male death rate increases. Since there are more susceptible females than males it implies that females are more likely to be infected and therefore males can afford to pay the cost of an increase to the infection rate. When male biased parasitism occurs it increases as harem size decreases. As harem size decreases the prevalence of infection decreases and the evolved level of transmission increases (disease resistance decreases) for both males and females (figure 3). This accentuates the relative difference between male and female transmission in a multiplicative manner.

A similar response occurs when male competition exceeds that of females (figure \ref{Fig4coevolpoint_qmbm}b) and can again be attributed to changes in the proportion of female/male susceptibles as the male competition parameter increases. When the male competition parameter is reduced below that of the female parameter the evolved level of transmission remains relatively constant. This occurs as the prevalence of infection also remains relatively constant and therefore there is no selection for a change in resistance to infection.

\subsection{Generality of results for other trade-offs}

We have undertaken the above analysis when the level of parasitism, $\beta$, is traded-off against the competition parameter, $q (q=g(\beta))$, and where we have imposed a trade-off such that a decrease in parasitism rate for females results in an increase in the parasitism rate for males ($\beta_m=g(\beta_f)$). We find the results are analogous to those presented above. The changes in the mating system do not produce a bias in parasitism between males and females. Male biased parasitism occurs when the males have a short lifespan (or where appropriate suffer increased competition) in comparison to females. Again the type of mating system can only accentuate this bias rather than cause it.

\section{Discussion}

Differences in the rate of parasitism between sexes and in particular male biased parasitism is often found in nature (see \citet{Skorping2004,Moore2002,Zuk1990,Zuk2009}). We examined whether parasite bias between the sexes could arise as a result of the mating system or through differences in the underlying life history characteristics between males and females through natural selection due to the epidemiological feedbacks that they cause. As such we are examining the evolutionary ecological implications of life-history and mating system in isolation from the role that they may play in sexual selection. Male-biased parasitism was selected for when males have a shorter lifespan than females or when males were subject to greater competition for resources than females (provided the overall level of competition was not too low). Female-biased parasitism will evolve if these conditions were reversed. Changes to the mating system did not produce a bias in parasitism but could accentuate an existing bias. In particular as harem size decreases an existing male biased parasitism is increased as a result of a decrease in overall prevalence. We therefore predict more male biased parasitism when males have shorter lifespans than females in monogamous or polyandrous species.

Selection for biased rates of parasitism requires there to be underlying differences in the life history characteristics of males and females. Male-biased mortality rates have been reported in vertebrate and invertebrate systems \citep{Promislow1992,Rolff2002} and it has been shown to have a positive correlation with male-biased parasitism \citep{Moore2002}. It has therefore been suggested that male-biased parasitism may drive the increased mortality rates in males \citep{Moore2002}. Our study indicates that male-biased parasitism may evolve as a consequence of male-biased mortality that was of course recognised as a possible interpretation of the empirical findings in \citet{Moore2002}. We show that the increased mortality in males leads to differences in the population dynamics with a greater proportion of susceptible females than males. This reduces the likelihood of infection for males and so they can afford to select for higher rates of parasitism. Fundamentally our argument is that non-disease causes of higher mortality in males \emph{per se} may select for the observed decrease in immune investment. This increase in mortality can come from processes such as increased risks from larger range size or fighting over females. By examining how natural selection operates through the ecological feedbacks we show in general terms that if males are shorter lived they will invest less in resistance.

The mating system, determined by the choice of the harem size, does not directly select for differences in parasitism between males and females in our models. However, differences in life-history characteristics that select for parasitism bias can be accentuated by the mating system. This is because as the harem size reduces, total prevalence levels also reduce leading to selection for reduced levels of host resistance. The mating system, which in this study acts via the harmonic birth function, effects population density and the overall level of disease resistance that evolves. If there is less selection for resistance then given an existing bias, it becomes accentuated. Taken as a whole, our results on the importance of mating systems are very different to those expected from sexual selection \citep{Zuk1990,Zuk2009,Stoehr2006,Moore2002}.  Generally polygamous species are expected to have stronger selection and therefore more male biased parasitism. Our results show that for a 50-50 sex ratio the ecological feedbacks that operate due to natural selection affect males and females equally. Therefore, the mating system \emph{per se} does not lead to a sexual bias. Furthermore, the effect of mating system on accentuating existing biases runs counter to the prediction of the sexual selection idea. Monogamous or polyandrous mating systems are more likely to accentuate the bias and therefore show male biased parasitism.

The mating system can have important consequences for the population dynamics that are exhibited and can lead to complicated (cycles, chaos) dynamical behaviour \citep{LINDSROM1998,MILLER2007a}. Recently, the population dynamical effects of male-biased parasitism for different case mortalities and both monogamous and polygamous mating systems \citep{MILLER2007a} have been examined. The population dynamics exhibited (point stability, cycles, chaos) did not show clear trends with increasing male-biased parasitism and the outcome depended on a complex interaction between the hosts mating system, demography and parasite virulence \citep{MILLER2007a}. Here we focus on the situation where the underlying population dynamics are at a stable point equilibrium to allow analysis of the fitness expression. We also choose the trade-off to ensure the singular point is an evolutionary stable attractor. Studies which examine the evolutionary behaviour for non-equilibrium underlying dynamics are rare but report that the evolutionary behaviour would not change for trade-offs where an attractor is predicted under equilibrium conditions \citep{WHITE2006,Hoyle2010}. We would therefore expect our finding to extend to non-equilibrium underlying dynamics.

Our results confirm previous general work on  the evolution of resistance to parasites \citep{Miller2007b,BOOTS2009} in that we find that the evolved level of host resistance increases as the average lifespan of the host increases. This result is linked to increased levels of prevalence which occur as lifespan increases. In our study the prevalence levels also increase as the harem size increases or as the level of competition for resources is reduced (until the competition for resources is low) (figure \ref{Fig1density}). As prevalence of infection increases it is beneficial to evolve higher levels of resistance in an attempt to avoid infection. Throughout we have assumed that there is no long-lived immunity after recovery from infection. In principle this may have important consequences with circumstances under which longer-lived individuals do not invest more in immunity. As such male biased parasitism may be less likely in disease with long lasting immunity, but a full theoretical analysis has not as yet been carried out.

This theoretical study examines the evolution of male-biased parasitism in the context of the complex epidemiological feedbacks in disease systems. A recent paper has also shown the importance of epidemiological feedbacks to the evolution of male biased parasitism \citep{Restif2010}. In a comprehensive study, they examine how differential exposure between males and females affects various aspects of investment in immunity under a range of trade-offs including one between recovery and lifespan \citep{Restif2010}. The model includes diploid genetics mapped onto a quantitative trait and fundamentally includes the epidemiological feedbacks caused by different investments under monogamous mating systems. \citet{Restif2010} show that reduced investment in males can evolve when there is more exposure to parasites (achieved by imposing differences in some of the disease characteristics between males and females). Their results further emphasise the importance of epidemiological feedbacks. Our study does not impose underlying differences in disease characteristics but focuses on the role of host life history and mating system. We have shown that male-biased parasitism can evolve as a consequence of sexual differences in life history characteristics that produce a greater proportion of susceptible females than males. Our results extend to different choices of trade-offs. Future studies should extend the analysis to examine the importance of the choice of underlying infectious disease framework and the representation of the two-sex birth function that may include assessing the effects of a non-equal sex ratio. Throughout we are focussing on the role of natural selection in the context of epidemiological feedbacks. Future work could combine this approach with models of sexual selection in order to gain a full understanding of the mechanisms that underpin male biased parasitism. The combined genetic and quantitative trait model of \citet{Restif2010} could be extended to provide a framework in which to examine these different processes.

\section{Acknowledgment}
Flora S.Bacelar acknowledges support from the Balaric Government, and from Spanish MICINN and FEDER through project FISICOS (FIS2007–60327) and to Em{\'\i}lio Hern{\'a}ndez-Garc{\'\i}a for reading the article and useful discussions. Andrew White is supported by a Royal Society of Edinburgh and Scottish Government Support Research Fellowship. Mike Boots is supported by Leverhulme Trust Fellowship




\newpage






\clearpage    
\section*{Figures}

\end{document}